# Ions-induced Epitaxial Growth of Perovskite Nanocomposites for Highly Efficient Light-Emitting Diodes with EQE Exceeding 30%

Zhaohui Xing[#], Guangrong Jin[#], Qing Du[#], Peiyuan Pang[#], Tanghao Liu, Yang Shen, Dengliang Zhang, Bufan Yu, Yue Liang, Dezhi Yang, Jianxin Tang, Lei Wang, Guichuang Xing*, Jiangshan Chen* and Dongge Ma*

Z. H. Xing, G. R. Jin, Q. Du, D. L. Zhang, B. F. Yu, Y. Liang, Prof. D. Z. Yang, Prof. J. S. Chen, Prof. D. G. Ma
Institute of Polymer Optoelectronic Materials and Devices, State Key Laboratory of Luminescent Materials and Devices, Guangdong Provincial Key Laboratory of Luminescence from Molecular Aggregates, South China University of Technology, Guangzhou 510640, China
Email: msjschen@scut.edu.cn (J. C.); msdgma@scut.edu.cn (D. M.)

P. Y. Pang, T. Liu, Prof. G. C. Xing
Joint Key Laboratory of the Ministry of Education, Institute of Applied Physics and Materials Engineering, University of Macau, Macau 999078, China
Email: gcxing@um.edu.mo (G. X.)

Dr. Y. Shen, Prof. J. X. Tang
Macao Institute of Materials Science and Engineering (MIMSE), Faculty of Innovation Engineering, Macau University of Science and Technology, Taipa 999078, Macau, China

Prof. L. Wang
Wuhan National Laboratory for Optoelectronics, Huazhong University of Science and Technology, Wuhan 430074, China

[#] These authors contribute equally to this research

**Abstract**

Metal halide perovskites, a class of cost-effective semiconductor materials, are of great interest for modern and upcoming display technologies that prioritize the light-emitting diodes (LEDs) with high efficiency and excellent color purity. The prevailing approach to achieving efficient luminescence from pervoskites is enhancing exciton binding effect and confining carriers by reducing their dimensionality or grain size. However, splitting pervoskite lattice into smaller ones generates abundant boundaries in solid films and results in more surface trap states, needing exact passivation to suppress trap-assisted nonradiative losses. Here, an ions-induced heteroepitaxial growth method is employed to assembe perovskite lattices with different structures into large-sized grains to produce lattice-anchored nanocomposites for efficient LEDs with high color purity. This approach enables the nanocomposite thin films, composed of three-dimensional (3D) $CsPbBr_3$ and its variant of zero-dimensional (0D) $Cs_4PbBr_6$, to feature significant low trap-assisted nonradiative recombination, enhanced light out-coupling with a corrugated surface, and well-balanced charge carrier transport. Based on the resultant 3D/0D perovskite nanocomposites, we demonstrate the perovskite LEDs achieving an remarkable external quantum efficiency of 31.0% at the emission peak of 521 nm with a narrow full width at half-maximum of only 18 nm. This research introduces a novel approach to the development of well-assembled nanocomposites for perovskite LEDs, demonstrating high efficiency comparable to that of state-of-the-art organic LEDs.

**Introduction**

Metal halide perovskites, valued for their low-cost solution processing, tunable optical bandgap, narrow emission bandwidth, strong defect tolerance and high charge carrier mobility,[1-3] hold a great promise for the development of high-performance photovoltaic and light-emission applications.[4-6] In recent years, researchers have dedicated substantial efforts to refine perovskites as efficient luminescent materials, focusing on achieving high photoluminescence quantum yields (PLQYs).[7-9] Moreover, perovskite light-emitting diodes (PeLEDs) have garnered significant attention, experiencing rapid advancements in enhancing the external quantum efficiencies (EQEs). This progress is attributed to their utilization of conventional device structures commonly found in organic light-emitting diodes (OLEDs). However, the EQEs of PeLEDs still fall short of those achieved by OLEDs. This discrepancy is partially attributed to the higher refractive index of perovskite,[10] resulting in inferior light out-coupling efficiency.[11-13] On the other hand, the performance of PeLEDs is limited by nonradiative losses including the trap-assisted nonradiative recombination. To propel efficiency of PeLEDs further, we must concurrently address three formidable requirements: high PLQY, high light extraction efficiency, and balanced carrier transport.

The progress in developing high-performance green LEDs holds significant significance, especially given the crucial role of green light as one of the primary colors essential for displays. $CsPbBr_3$ emerges as a prominently used inorganic emitter in green PeLEDs, capitalizing on its remarkable thermal and chemical stability when compared to organic-inorganic hybrid perovskites. Despite these advantages, its low exciton binding energy contributes to suboptimal exciton recombination. To integrate $CsPbBr_3$ into PeLEDs and achieve high efficiency, a deliberate enhancement of its exciton binding effect becomes imperative. The common way for improving exciton binding energy of pervoskites is to split their crystal lattices into small ones by reducing their dimensionality or size to induce quantum confinement effect.[14,15] Previously, we incorporated excess CsBr into the saturated precursor solution to derease the size of $CsPbBr_3$ nanocrystals in the perovskite emissive layers (EMLs), leading to an increased exciton binding energy.[16] In addition, the formation of $Cs_4PbBr_6$ phase was aroused after introducing excess CsBr. It is notable that $Cs_4PbBr_6$ possesses a lower refractive index compared to $CsPbBr_3$, allowing $Cs_4PbBr_6$ grains to function as low-index grids in the EMLs.[17,18] This arrangement facilitates the escape of wide-angle light trapped within PeLEDs.[19,20] However, the reduction in crystal size leads to an increased surface area of $CsPbBr_3$, resulting in a higher density of surface defects. Despite $Cs_4PbBr_6$ being capable of passivating the surface defects of $CsPbBr_3$ due to their closely matched lattice structures,[21] the PLQYs of the perovskite films, which are positively correlated to the EQEs of PeLEDs, remains relatively low. Therefore, a novel strategy must be devised to enhance the efficiency of $Cs_4PbBr_6$ in interacting with the lattice on the surface of $CsPbBr_3$, aiming to improve the overall PLQYs of the perovskite films for high efficiency PeLEDs.

Here in, we developed an innovative ions-induced epitaxial growth method to synthesize large-sized $CsPbBr_3$&$Cs_4PbBr_6$ nanocomposites in which $Cs_4PbBr_6$ strongly passivated the embedded $CsPbBr_3$ emitters. Sodium ions ($Na^+$) were incorporated as a 'coordinator' to encourage the heteroepitaxial growth between $CsPbBr_3$ and $Cs_4PbBr_6$, forming coarsened nanocomposite grains with lowered trap states. The grain sizes of the $CsPbBr_3$&$Cs_4PbBr_6$ nanocomposites greatly exceed the average thickness of the perovskite film, resulting in a concave-convex morphological structure facilitating efficient light extraction. Furthermore, we addressed the crucial issue of carrier transport imbalance, essential for high-performance PeLEDs[22,23], by incorporating formamidinium chloride (FACl) to enhance electron transportation. Ultimately, we successfully demonstrated the PeLEDs with the $CsPbBr_3$&$Cs_4PbBr_6$ nanocomposite films as the EMLs, featuring significantly low trap density, improved light extraction efficiency and enhanced carrier balance. The champion device achieved a remarkable EQE of 31.0% and a maximum current efficiency of 113.4 cd $A^{-1}$ at the emission peak of 521 nm with a narrow full width at half-maximum (FWHM) of 18 nm.

**Results and Discussion**

In this study, our objective was to assemble $CsPbBr_3$ with $Cs_4PbBr_6$ to create well-passivated luminescent $CsPbBr_3$&$Cs_4PbBr_6$ nanocomposites. The perovskite films were prepared by a one-step spin-coating method, eliminating the need for antisolvent treatment.[16,24,25] The pristine precursor consisted of phenethylammonium bromide (PEABr), CsBr and $PbBr_2$ in a ratio of 0.4:1.6:1. The films and devices based on this pristine precursor are hereafter denoted as "w/o". Additionally, NaBr and formamidinium chloride (FACl) were introduced into the pristine precursor as additives, with each accounting for 10% of the $PbBr_2$ content. And the samples with only NaBr and both of NaBr and FACl are denoted as "w/NaBr" and "w/NaBr:FACl", respectively. **Fig. 1a** shows the absorption and photoluminescence (PL) spectra of the w/o, w/NaBr and w/NaBr:FACl films. As expected, the absorption edges and peaks at around 515 nm and 313 nm, which are respectively ascribed to the $CsPbBr_3$ and $Cs_4PbBr_6$ phases, are obviously found in all the three films. And the absorption peaks corresponding to other low-dimensional phases are very weak. These absorption results suggest that the major components of the as-prepared perovskite films are $CsPbBr_3$ and $Cs_4PbBr_6$ without or with additional ions from NaBr and FACl. The existence of $Na^+$, $FA^+$ and $Cl^-$ are confirmed by the X-ray photoelectron spectroscopy (XPS) and secondary ion mass spectrometry (SIMS), as shown in **Fig. S1** and **Fig. S2**. The PL peak of the w/o film is located at 516 nm, which belongs to the emission from the $CsPbBr_3$ phase. After the addition of NaBr, there is no change in the PL spectrum, as well as in the absorption edge, suggesting that the $Na^+$ ions do not affect the structure and bandgap of the $CsPbBr_3$ emitter in the w/NaBr film. However, it is observed that the PL peak of the w/NaBr:FACl film is red-shifted to 520 nm, being consistent with the change of the absorption edge. This change should be attributed to the larger cation effect, leading to octahedral expansion after incorporating $FA^+$ ions into $CsPbBr_3$.[26]

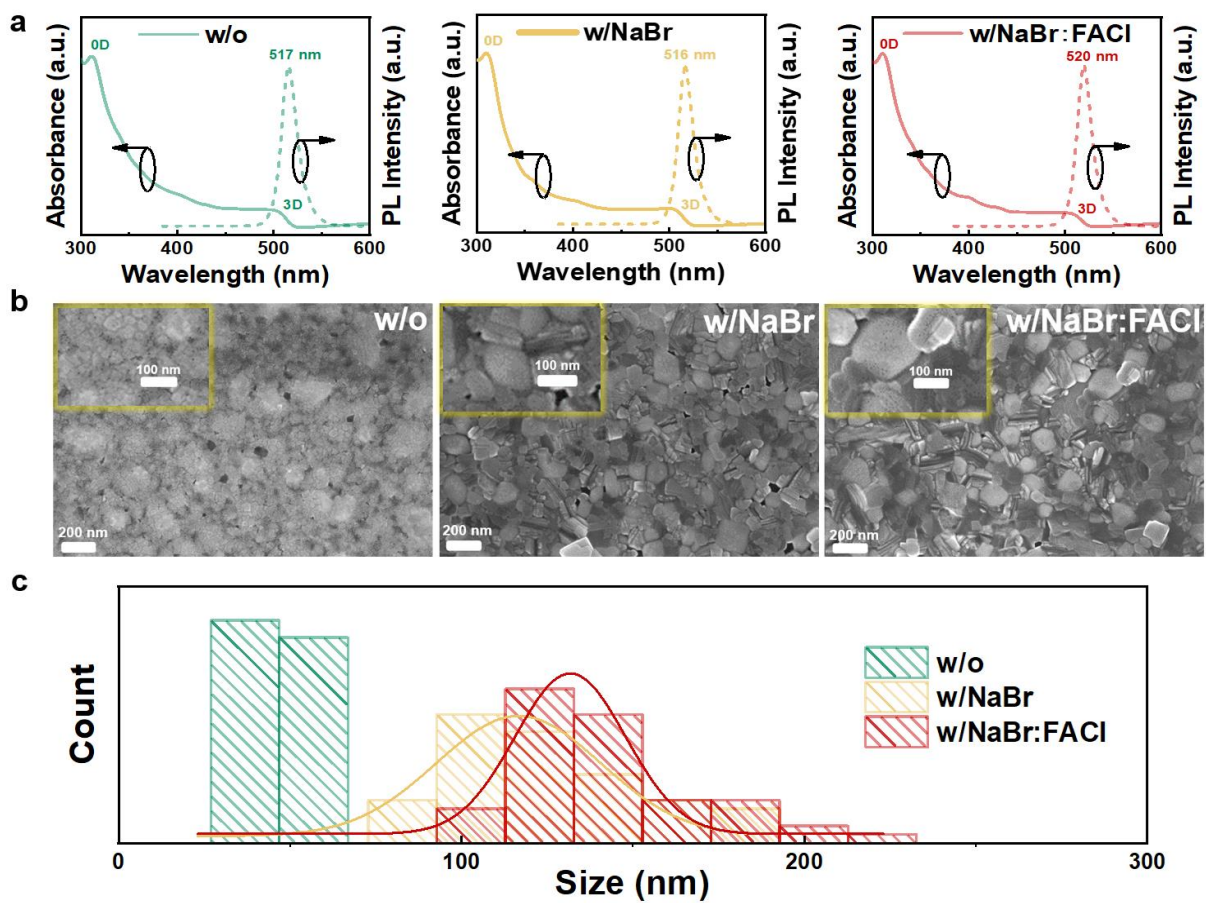

**Figure 1** (a) The absorption and PL spectra of the w/o, w/NaBr and w/NaBr:FACl films indicate that the primary constituents of the perovskite films, as prepared, are $CsPbBr_3$ and $Cs_4PbBr_6$. This holds true whether or not additional ions from NaBr and FACl are introduced. (b) SEM images depicting the perovskite films of w/o, w/NaBr and w/NaBr:FACl reveal that large sized grains become prominent in the latter two cases. (c) The distribution of grain size in the perovskite films of w/o, w/NaBr and w/NaBr:FACl suggests a noticeable increase in grain size with additives.

To explore the effect of additives on the structural characteristics of the perovskite samples, the film morphology was studied by using scanning electron microscopy (SEM). Notably, the SEM images of the perovskite films, as depicted in **Fig. 1b**, show substantial disparity after incorporating NaBr and FACl. It is obvious that large sized grains become prominent in the w/NaBr and w/NaBr:FACl films, implying that grain coarsening is occurred. The particle size distribution for each sample is presented in **Fig. 1c**. The average size of grains in the w/NaBr film is significantly increased, indicating the crystal growth is strongly affected by the addition of NaBr. The grains become larger after further adding FACl, leading to a little increase of average size and dense packing in the w/NaBr:FACl film. These results indicate that the grain coarsening effect is induced by sodium ion ($Na^+$) and can be intensified by adding FACl.

Based on the above absorption and SEM results, we speculate that the grain coarsening should be correlative with the growth of $CsPbBr_3$ and $Cs_4PbBr_6$ during the film formation. To get more insight into the grain coarsening initialized by $Na^+$, we compared the scanning transmission electron microscopy (STEM) images of the perovskites before and after adding NaBr, as shown in **Fig. 2a** and **b**. The samples for STEM measurements were prepared by peeling off the as-fabricated films and dispersing in chlorobenzene, then dropping on copper mesh and drying in a vacuum. It is obvious that the w/o and w/NaBr samples show much different grain size and shape, which is consistent with the SEM results. In the pristine sample, there are a great number of angular grains with small size. After the addition of $Na^+$, there emerge larger grains with irregular shapes. To analyze the species of the large grains in the w/NaBr sample, the high-resolution transmission electron microscopy (HRTEM) measurement was carried out. The HRTEM image was illustrated in the insert of **Fig. 2b**. The characteristic lattice spacings of 0.41 nm and 0.69 nm should belong to the $CsPbBr_3$ (110) and $Cs_4PbBr_6$ (110) lattice planes, respectively. This finding reveals that the large grains consist of a dual-phase composition of $CsPbBr_3$ & $Cs_4PbBr_6$.

Subsequently, we used the energy dispersive X-ray spectroscopy (EDS) to assess the elements composition at different positions of the $CsPbBr_3$&$Cs_4PbBr_6$ nanocomposites, as shown in **Fig.S3**. The analysis revealed a notably higher concentration of Cs and Br within the large grains. To be more specific, the small and bright spots corresponding to the '001' region exhibited a Cs:Pb:Br ratio of 1.86:1:6.84, which closely resembles the composition with high proportion of $CsPbBr_3$. In contrast, the dark matrix surrounding these spots, referred to as '002', exhibited a Cs:Pb:Br ratio of 2.88:1:8.24, resembling the composition with high proportion of $Cs_4PbBr_6$. This observation suggests that the $Cs_4PbBr_6$ tends to envelop the $CsPbBr_3$ in the large sized CsPbBr$_3$&Cs$_4$PbBr$_6$ nanocomposites, effectively shielding CsPbBr$_3$ from boundary defects.[21]

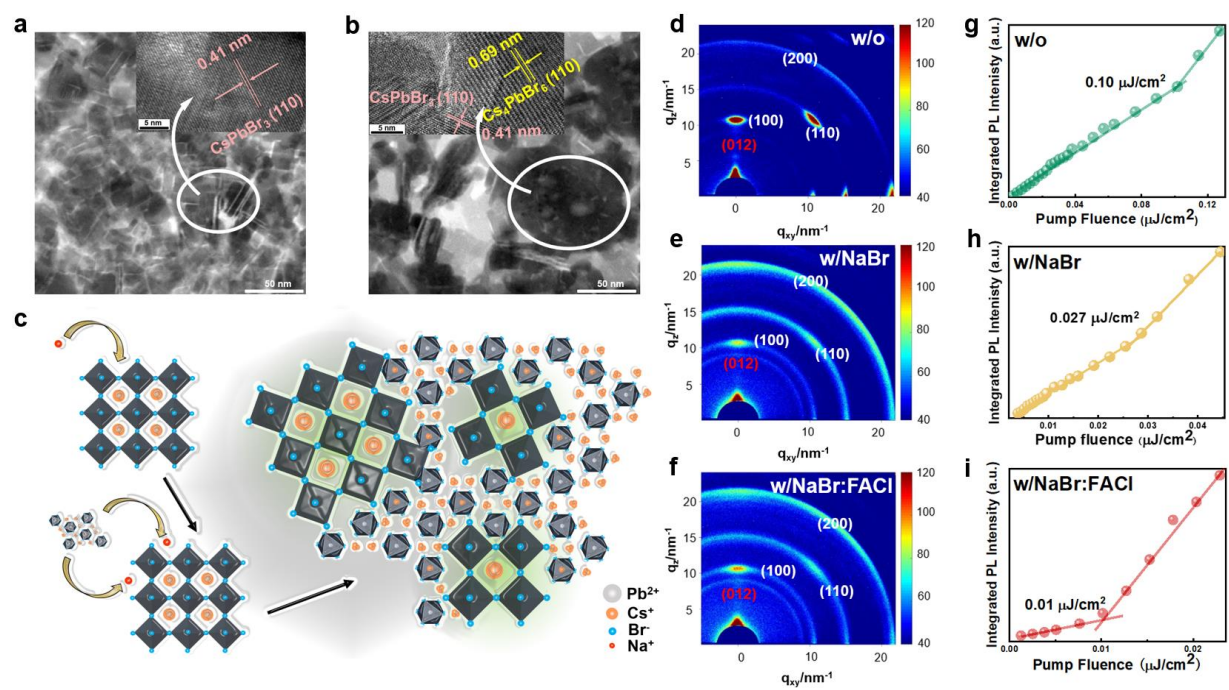

**Figure 2** The STEM and corresponding HRTEM (inserted) patterns of (a) w/o perovskite film, and (b) w/NaBr perovskite film. (c) Schematic diagram of the formation processes of large sized CsPbBr$_3$&Cs$_4$PbBr$_6$ nanocomposites induced by Na$^+$ ions. Na$^+$ ions initially adhere to the surface of CsPbBr$_3$ and then promote the epitaxial growth of Cs$_4$PbBr$_6$ from the surface of CsPbBr$_3$. (d-f) The GIWAXS patterns of the w/o, w/NaBr and w/NaBr:FACl perovskite films, respectively. (g-i) The threshold pump fluence values for the perovskite films of w/o, w/NaBr and w/NaBr:FACl.

To further understand the coarsening mechanism of CsPbBr$_3$&Cs$_4$PbBr$_6$ induced by Na$^+$, we carried out the density functional theory (DFT) simulations, detailed in Supporting Information. The calculated formation energy (listed in **Table S1**) suggests that the Na$^+$ has limited affinity for occupying the A site within the CsPbBr$_3$ lattice, which can be validated by the unchanged bandgap (according to the unchanged PL peaks and absorption edges in **Fig. 1a**) and the obtained (110) lattice spacing (**Fig. 2a**). Instead, Na$^+$ tends to attach on the surface of CsPbBr$_3$, as shown in **Fig. S4**. This attachment is thermodynamically favored, as corroborated by the adsorption energy as listed in **Table S2**. Furthermore, Na$^+$ reduces the formation energy of the Cs$_4$PbBr$_6$ phase, as evidenced in **Table S3**. These results suggest that the introduced Na$^+$ ions would initially approach to the surface of CsPbBr$_3$ and then facilitate the epitaxial growth of Cs$_4$PbBr$_6$, resulting in the formation of large grains with CsPbBr$_3$ embedding into Cs$_4$PbBr$_6$ matrix. The schematic diagram of formation processes of the CsPbBr$_3$&Cs$_4$PbBr$_6$ nanocomposites is illustrated in **Fig. 2c**.

To elucidate the overall crystal structure of the perovskite films, we performed the X-ray diffraction (XRD) measurements, as presented in **Fig. S5**. The diffraction peaks at 2θ values of 15.2° and 30.4° correspond to the lattice planes of (100) and (200) for Pm-3m cubic CsPbBr$_3$ structure. The pristine film displays more pronounced diffraction peaks compared to the w/NaBr and w/NaBr:FACl films, reflecting that the ions-induced growth of Cs$_4$PbBr$_6$ can sufficiently confine the growth of CsPbBr$_3$. The lowered crystallinity also indicates that the alignment of CsPbBr$_3$ phase could become less ordered in the large CsPbBr$_3$&Cs$_4$PbBr$_6$ nanocomposites. In order to investigate the effect of additives on the crystal orientation characteristics of the perovskite films, grazing-incidence wide-angle X-ray scattering (GIWAXS) measurements were conducted. **Fig. 2d-f** show the GIWAXS patterns of the w/o, w/NaBr and w/NaBr:FACl films for comparison. The scattering rings at q = 10.7, 15.3 and 21.4 nm$^{-1}$ correspond to the (100), (110) and (200) crystallographic planes of CsPbBr$_3$, respectively. It is clearly seen that the w/o film presents typical discrete scattering spots in the GIWAXS pattern, indicative of the formation of highly oriented CsPbBr$_3$ phases. In contrast, the w/NaBr and w/NaBr:FACl films show less ordered orientation of the CsPbBr$_3$ phases, as demonstrated by the significantly weakened scattering spots and enhanced isotropic scattering rings. In addition, the scattering ring at around q = 9.0 nm$^{-1}$, which is assigned to the (012) plane of Cs$_4$PbBr$_6$, was detected in all the three films. The integration intensity profiles of the scattering rings, as displayed in **Fig. S6**, reveals the enhanced crystallinity of Cs$_4$PbBr$_6$ in the w/NaBr and w/NaBr:FACl films. These GIWAXS results clearly demonstrated that the incorporation of NaBr and FACl additives resulted in the significant changes of the orientation and crystallinity of $CsPbBr_3$ and $Cs_4PbBr_6$ in the coarsened $CsPbBr_3$&$Cs_4PbBr_6$ nanocomposite films, where $CsPbBr_3$ is randomly located in the $Cs_4PbBr_6$ matrix as illustrated in **Fig. 2c**.

It is noticed that the proportion of large grains get increased after adding FACl (**Fig. 1c**). This is attributed to the decreased formation energy of $Cs_4PbBr_6$ with FACl (as shown in **Table S4**), benefiting to the growth of $Cs_4PbBr_6$ for coarsening $CsPbBr_3$&$Cs_4PbBr_6$ nanocomposites. The promoting growth of $Cs_4PbBr_6$ should result from the synergistic effect of $FA^+$ and $Cl^-$ because adding $FA^+$ alone is adverse to the formation of $Cs_4PbBr_6$ phase as shown in **Table S5**. The increased proportion of large $CsPbBr_3$&$Cs_4PbBr_6$ nanocomposites leads to $Cs_4PbBr_6$ more sufficiently encapsulating $CsPbBr_3$, indicating more effective passivation of $CsPbBr_3$ by $Cs_4PbBr_6$. In addition, the $Cl^-$ anions not only assist in the formation of $Cs_4PbBr_6$ phase (**Table S5**), but also have the capability to occupy halogen vacancies within $CsPbBr_3$[27,28]. DFT simulations demonstrate that the $Cl^-$ is more likely to occupy the halogen positions in $CsPbBr_3$ than $Br^-$, as shown in **Table S6**. This property allows the $Cl^-$ ions to act as an effective passivation agent to reduce the ionic defects of halogen vacancies. Furthermore, the formation energy of the $CsPbBr_3$ with both of $Cl^-$ and $FA^+$ is much lower than those of the $CsPbBr_3$ with $Cl^-$ and $FA^+$ alone (**Table S6**), confirming the codoping of $Cl^-$ and $FA^+$ ions. Based on the above discussions, it can be proposed that the nanocomposite grains in the w/NaBr:FACl film are composed of the $CsPbBr_3$ phase incorporating with $FA^+$ and $Cl^-$ and the $Cs_4PbBr_6$ phase with $Na^+$, $FA^+$ and $Cl^-$, which lead to an observable change in the Pb 4f peaks (see the XPS patterns in **Fig. S1c**). Thanks to the passivation of $CsPbBr_3$ by $Cs_4PbBr_6$ and $Cl^-$, the trap densities in the perovskite films are dramatically reduced from $3.08 \times 10^{15}$ cm$^{-3}$ (w/o) to $5.92 \times 10^{14}$ cm$^{-3}$ (w/NaBr) and $2.69 \times 10^{14}$ cm$^{-3}$ (w/NaBr:FACl), as seen in **Table S7**, along with a reduced threshold pump fluence ($P_{th}^{trap}$) at which the trap states become fully occupied[6,29] (**Fig. 2g-i**). As a result, the PLQYs of the perovskite films are significantly improved from 44.8% (w/o) to 81.4% (w/NaBr) and 79.2% (w/NaBr:FACl) (**Fig. S7**). Generally, the low trap density and high PLQY would be conducive to achieve efficient PeLEDs.

To investigate the influence of adding NaBr and FACl on the EL characteristics of the perovskite nanocomposite films, we fabricated the PeLEDs with a structure of indium tin oxide (ITO)/nickel oxide (NiO$_x$)/poly[bis(4-phenyl) (2,4,6-trimethylphenyl)amine] (PTAA)/poly(9-vinylcarbazole) (PVK)/perovskite/1,3,5-Tris(1-phenyl-1H-benzimidazol-2-yl)benzene (TPBi)/LiF/Al, as depicted in **Fig. 3a**, The performances of the w/o, w/NaBr and w/NaBr:FACl devices are shown in **Fig. 3b-f**. Compared to the control device of w/o, the devices of w/NaBr and w/NaBr:FACl show higher current density and luminance under the same operating voltage (**Fig. 3b**). As predicted, the w/NaBr and w/NaBr:FACl devices exhibit improved EL efficiency (**Fig. 3c** and **d**). It is noteworthy that the maximum EQE of 31.0% was achieved in the champion device of w/NaBr:FACl. Correspondingly, the champion w/NaBr:FACl device possesses a high current efficiency (CE) of 113.4 cd A$^{-1}$. From the EL spectra as shown in **Fig. 3e**, it can be found that the w/o and w/NaBr devices present unchanged EL peaks at 517 nm, but the w/NaBr:FACl device shows an EL peak red-shifted to 521 nm, which is in good agreement with the change of the PL spectra as discussed above (**Fig. 1a-c**). Additionally, a narrow FWHM of 18 nm is obtained from the EL spectra of the PeLEDs, representing the green emission with high color purity. To our knowledge, the resultant w/NaBr:FACl device is among the highly efficient green PeLEDs reported to date, and is the only one with the EQE surpassing 30% and the FWHM less than 20 nm (see **Table S8**, Supporting Information). To make a better comparison, the statistic maximum EQEs are given in **Fig. 3f**.

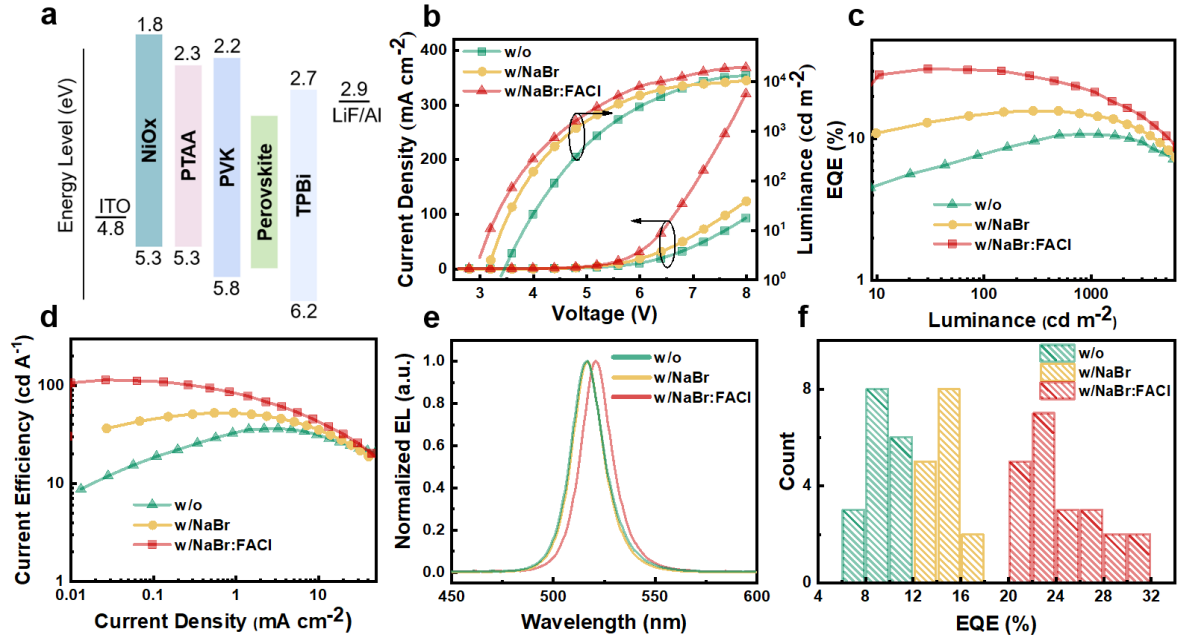

**Figure 3** (a) Diagram of the used device structure. (b) Current-Voltage-Luminance characteristics. (c) EQE-Luminance characteristics. (d) Current Efficiency-Current Density characteristics. (e) EL spectra of the devices under an applied voltage of 5V. (f) Histogram of peak EQEs of the devices.

Besides lowering trap density by passivation, the formed concave-convex structure should play an important role in the performance improvement of the PeLEDs because it can enhance the light extraction efficiency. To analyze the effect of the concave-convex structure on the light extraction, we conducted atomic force microscope (AFM) measurements to get more characteristics about the spatial morphology of the perovskite films (**Fig. 4a**, with three dimensional images in **Fig. S8**). Given that the average thickness of the perovskite layer is only around 45 nm, significantly smaller than the average size of the large grains, resulting in a concave-convex structure. And then, we used cross-section TEM to further examine the concave-convex structure caused by grain coarsening. The w/o sample reveals a homogeneous and flat film (**Fig. 4b**) for comparation. We prepared the cross-sectional TEM sample of w/NaBr:FACl by using the complete device coated with C/Pt (**Fig. 4c**, with elements mapping in **Fig. S9**). The influence of the large $CsPbBr_3$&$Cs_4PbBr_6$ nanocomposites on spatial morphology is notable, introducing an undulating EML. Furthermore, the morphology of the perovskite layer impacts those of the vacuum-evaporated TPBi and Al layers above, which significantly contributes to optical outcoupling.[30] Meanwhile, the $Cs_4PbBr_6$ grains can play the role of low-index grids between the $CsPbBr_3$ grains, helping wide-angle light trapped coming out of the PeLEDs.[19,20]

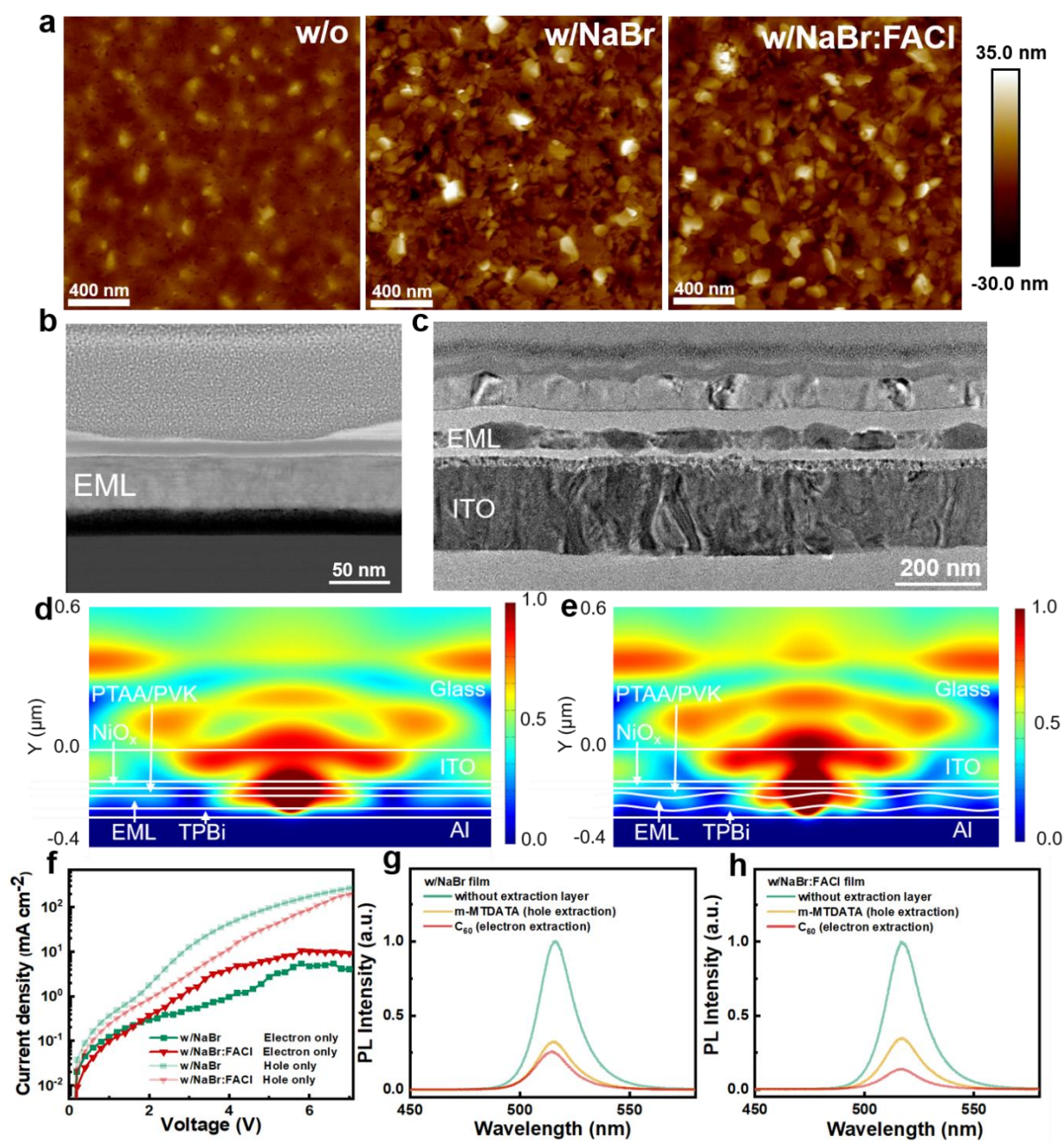

**Figure 4** (a) AFM images of the w/o, w/NaBr and w/NaBr:FACl perovskite films. (b) Cross-section TEM image of the w/o sample. (c) Cross-section TEM image of the w/NaBr:FACl sample in the device structure. (d-e) Stimulated near-field transverse electric (TE) distributions for the w/o and w/NaBr:FACl PeLEDs, respectively. (f) Current density-Voltage characteristics of the electron-only and hole-only devices based on the w/NaBr and w/NaBr:FACl films. (g-h) Quenching effects in the w/NaBr and w/NaBr:FACl films, respectively.

To assess the contribution of large $CsPbBr_3$&$Cs_4PbBr_6$ nanocomposites to the light out-coupling efficiency, we conducted optical models using the finite-difference-time-domain (FDTD) approach. Randomly oriented dipoles placed at different positions were used to stimulate the near-field transverse electric (TE) distribution. The device structure of PeLEDs consisted of ITO (185 nm)/$NiO_x$ (20 nm)/PTAA&PVK (25 nm)/perovskite EML (45 nm)/TPBi (40 nm)/LiF (1 nm)/Al (100 nm). The flat model is depicted in **Fig. 4d** for comparison. Pyramid model was used to represent perovskite grains, roughly corresponding to the AFM images, featuring a periodicity of 200 nm with a height of 20 nm for the w/NaBr:FACl EML (**Fig. 4e**). Moreover, the TPBi layer and Al electrode exhibit half of the bending observed in the perovskite layer, according to the cross-section TEM. Consequently, the presence of large grains and corresponding bent electrode result in enhanced outcoupling efficiency compared to a flat device structure. The extent of light extraction efficiency enhancement is dependent on the repetition frequency and height of the pyramid model. The increased proportion of large $CsPbBr_3$&$Cs_4PbBr_6$ gains in the w/NaBr:FACl EML could contribute more light extraction in the PeLEDs.

As shown in **Fig. 3b**, the current density of the w/NaBr:FACl device is much higher than that of the w/o and w/NaBr devices, which could be another contributing factor to the more improvement of EL performance. The increased current density demonstrates that the addition of FACl is helpful to improve the electrical properties of the perovskite film. As known, the aforementioned dual-phase composition and trap states as well as the concave-convex morphology could influence the electrical properties including charge carrier injection and transportation. To understand the effect of adding FACl on the electrical properties, we fabricated the electron-only devices of ITO/ZnO/perovskite/TPBi/LiF/Al and the hole-only devices of ITO/$NiO_x$/PTAA/PVK/perovskite/4,4',4''-tris(N-carbazolyl) triphenylamine (TCTA)/$MoO_3$/Al. The current-voltage characteristics of these devices are shown in **Fig. 4f**. It can be seen that the electron current is increased while the hole current is decreased after incorporating FACl into the w/NaBr film, demonstrating the improved balance of carrier transportation which is essential to achieve efficient PeLEDs. In the $CsPbBr_3$&$Cs_4PbBr_6$ dual-phase films, the $Cs_4PbBr_6$ phase can be used to confine the excitons for radiative recombination, but it will inhibit carrier injection and transportation because of its large bandgap and low conductivity. The increased electron current in the electron-only device of w/NaBr:FACl would be partly due to the reduction of trap density for better electron transportation. On the other hand, the electron injection into the $CsPbBr_3$ emitter, which is limited by the surrounding $Cs_4PbBr_6$ barrier, would be improved after the addition of FACl. Therefore, we conducted PL-quenching experiments to investigate the effect of adding FACl on the carrier across the $Cs_4PbBr_6$ barrier. We introduced an electron-extraction layer of $C_{60}$, whose lowest unoccupied molecular orbital (LUMO) energy level is below the conduction band of the $CsPbBr_3$ phase in the perovskite EMLs, as well as a hole-extraction layer of 4,4',4''-Tris[phenyl(m-tolyl) amino] triphenylamine (m-MTDATA) with the highest occupied molecular orbital (HOMO) energy level above the valence band of the $CsPbBr_3$ phase. These carrier-extraction layers allow to decouple of the electron and hole dynamics within the perovskite materials.[31] **Fig. 4g-h** present the PL intensity of the w/NaBr and w/NaBr:FACl perovskite films covered by different carrier-extraction layers, in comparison to the bare perovskite films without a carrier-extraction layer, respectively. It can be found that the PL intensities are significantly reduced when covered with the electron-extracting $C_{60}$ layer or the hole-extracting m-MTDATA layer. Most notably, there is an obviously additional quenching effect observed in the w/NaBr:FACl EML covered by the $C_{60}$ layer, attributed to an enhanced capability for electron extraction, indicating improved electron across the $Cs_4PbBr_6$ barrier. On the contrast, the same quenching effect was observed in the w/NaBr and w/NaBr:FACl films covered by the m-MTDATA layer, suggesting the same capability for holes crossing the $Cs_4PbBr_6$ barrier. In short, the improved carrier balance in the w/NaBr:FACl PeLED can be attributed to the enhanced electron injection and transportation.

As discussed above, the greatly improved performance in the w/NaBr:FACl PeLED should be resulted from the synergistic effects of lowering trap density and enhancing light extraction efficiency, as well as improving carrier balance. Our study demonstrated that the construction of coarsened $CsPbBr_3$&$Cs_4PbBr_6$ nanocomposites by ions-induced epitaxial growth can provide an effective way for the development of highly efficient PeLEDs. In the future, more efforts will be made to improve the operational stability, which is the common challenge for PeLEDs.

**Conclusion**

In summary, we have demonstrated an ions-induced crystallization approach to synthesize the perovskite dual-phase nanocomposites of $CsPbBr_3$&$Cs_4PbBr_6$ with $CsPbBr_3$ effectively encapsulated by $Cs_4PbBr_6$. The ions-induced growth of perovskite nanocomposites enables the enlargement of gain sizes, leading to reducing trap density in the perovskite EMLs with well-passivated $CsPbBr_3$ emitter, enhancing light extraction efficiency in the concave-convex structural devices, and improving electrical characteristics with balanced carrier injection and transportation. As a result, the green PeLEDs with the maximum EQE exceeding 30% and a narrow FWHM of 18 nm were successfully achieved. We believe that this straightforward method of enlarging grain size to reduce trap density and enhance light out-coupling, along with an easy approach of adjusting the electrical properties of devices, will unlock new opportunities for high-performance PeLEDs.

## Acknowledgement

This work was supported by the National Key Research and Development Program of China (2022YFE0206000), the National Natural Science Foundation of China (U2001219, 51973064, 61935017), the Guangdong Basic and Applied Basic Research Foundation (2023B1515040003, 2019A1515012142), the Natural Science Foundation of Guangdong Province (2023B1212060003), the Science and Technology Development Fund (FDCT), Macau SAR (No. 0010/2022/AMJ, No. 0018/2022/A1), the Hubei Provincial Natural Science Foundation of China (2023AFA034, 2023BAB102), the Open Project Program of Wuhan National Laboratory for Optoelectronics (NO. 2021WNLOKF014), and the State Key Lab of Luminescent Materials and Devices, South China University of Technology (Skllmd-2023-05). The authors thank Prof. Yonghua Chen and Lingfeng Chao from Nanjing Tech University for the assistance with the GIWAXS measurements, thank Chengwei Lin, Ji Li and Sizhe Tao from South China University of Technology for the assistance with the PL-Quenching experiments.

## Conflict of interest

The authors declare no competing interest

## Methods

**Materials.** Nickel acetate tetrahydrate (Ni(CH$_3$COO)$_2$·4H$_2$O) (99.9%) and Ethanolamine (99%) were purchased from ACROS ORGANICS. PEABr (99.5%), FABr (99.5%), FACl (99.5%) and PTAA (average Mw = 6 000-15 000) were purchased from Xi'an Polymer Light Technology Corporation. PVK (average Mw = 1 100 000), PbBr$_2$ (99.999%), CsBr (99.999%), NaBr (99.5%), DMSO (anhydrous, 99.9%), ethanol (anhydrous) and chlorobenzene (anhydrous, 99.8%) were purchased from Sigma-Aldrich. TPBi, TCTA, C$_{60}$, m-MTDATA and LiF were purchased from Jilin OLED Photoelectric Material Corporation. All the materials were used without further purification.

**Preparation of perovskite precursor.** The perovskite precursor solutions were prepared by dissolving PEABr, CsBr, NaBr, FABr, FACl and PbBr$_2$ at the desired molar ratios in anhydrous DMSO. The concentration of PbBr$_2$ was set at 0.3 M. The solutions were stirred at 48 °C for 10 hours and then filtered with a polytetrafluoroethylene filter (0.22 μm) before use.

**Device fabrication and evaluation.** The devices were fabricated on 0.70 mm thick glass substrates precoated with an indium tin oxide (ITO) layer (185 nm) with a sheet resistance of 8 Ω per square. The ITO surface was cleaned with detergents and deionized water by ultrasonic, then dried at 120 °C for 1 h. The NiO$_x$ precursor was prepared using the previous method[32] and then spin-coated on ITO at 4000 r.p.m. for 30s followed by annealed at 350 °C for 60 min in air. Then, the substrates were transferred into a nitrogen-filled glove box (H$_2$O < 1 ppm, O$_2$ < 1 ppm) after cooling. The PTAA was spin-coated at 4000 r.p.m. from a chlorobenzene solution with a concentration of 8 mg mL$^{-1}$ and then dried at 170 °C for 30 min. After that, the PVK (8 mg mL$^{-1}$ in chlorobenzene) was spin-coated on the PTAA layer and then dried at 170 °C for 30 min. The perovskite EMLs were spin-coated at 4000 r.p.m for 120 s and then annealed at 110 °C for 10 min. Finally, the samples were transferred into a thermal evaporator where a TPBI layer (40 nm), a LiF layer (1 nm) and an Al layer (100 nm) were deposited layer by layer at a pressure below $10^{-4}$ Pa. The device active area was 8 mm$^2$, as defined by the overlapping area of the Al cathode and the ITO anode. We used the LED measurement method of Forrest et al.,[33] at room temperature under ambient atmosphere with encapsulation. The current density-luminance-voltage (J-L-V) characteristics were performed simultaneously using a computer-controlled source meter (Keithley 2400) equipped with a light intensity meter (Konica Minolta LS-110). The EL spectra were recorded with a spectrometer (Ocean Optics USB2000+). The EQEs were calculated from the luminance, current density, and EL spectra, assuming a Lambertian distribution (see **Fig. S10**). All results from the devices were measured in the forward-viewing direction.

**Characterizations.** The ultraviolet-visible absorption spectra were measured with a Shimadzu UV-3600 UV-vis-NIR spectrophotometer. The PL spectra were recorded using a Horiba Fluoromax-4 spectrofluorometer with an excitation wavelength of 365 nm. PLQYs of the perovskite films were measured using a commercialized PLQY measurement system from Ocean Optics with a 365 nm LED as the excitation light source. The broadband femtosecond TA spectra of the perovskite films were taken using an Ultrafast System HELIOS TA spectrometer. The laser source was a Coherent Legend regenerative amplifier (100 fs, 1 kHz, 400 nm) seeded by a Coherent Vitesse oscillator (100 fs, 80 MHz). The XRD characterization were performed with the Rigaku Smart lab (9 kW) with a Cu-K$_\alpha$ X-ray tube ($\lambda$ = 1.5405 Å). The GIWAXS measurement was performed on the BL14B1 beamline of the Synchrotron Radiation Facility (SSRF) in Shanghai, China. The SEM images were obtained by a HITACHI Regulus-8100. The AFM images were obtained with a Bruker Multimode 8. The HRTEM, STEM, SAED and EDS were conducted by a JEM-F200, with the sample being fragments of the emissive layer. The samples for cross-section TEM were prepared by the focusing ion beam (FIB) process, which conducted by Zeiss Crossbeam 540. The secondary ion mass spectrometry (SIMS) depth analysis was obtained by the IONTOF 5, with the sample being the complete device structure. The UPS and XPS diagrams were obtained by the Thermo SCIENTIFIC Nexsa.